# LINK SPAM DETECTION BASED ON DBSPAMCLUST WITH FUZZY C-MEANS CLUSTERING

Dr.S.K.Jayanthi[1] and Ms.S.Sasikala[2]

[1]Associate Professor & Head, Computer Science Department,
Vellalar college for women, Erode, India
`jayanthiskp@gmail.com`
[2]Lecturer, Computer Science Department,
KSR College of Arts and Science, Tiruchengode, India
`sasi_sss123@rediff.com`

## ABSTRACT

*This Search engine became omnipresent means for ingoing to the web. Spamming Search engine is the technique to deceiving the ranking in search engine and it inflates the ranking. Web spammers have taken advantage of the vulnerability of link based ranking algorithms by creating many artificial references or links in order to acquire higher-than-deserved ranking n search engines' results. Link based algorithms such as PageRank, HITS utilizes the structural details of the hyperlinks for ranking the content in the web. In this paper an algorithm DBSpamClust is proposed for link spam detection. As showing through experiments such a method can filter out web spam effectively.*

## 1. INTRODUCTION

The Web is potentially a terrific place to get information on almost any topic. Search engines play an important role in locating desired information from millions of web pages, and people become increasingly rely on the search results. Therefore, search engine have the vital influence in the visits of many websites, especially these sites which have highly rankings in the search result can get high visits easily.

Nowadays, link-based ranking algorithms like PageRank, HITS, rank every web page based on both the number of incoming links a web page has and the weight of these incoming links. An increase in ranking list requires a large number of incoming links from low-PageRank pages and/or some hard-won links from well known websites. It is obvious for spammers that the latter requirement is not as feasible as the former to reach, on account of the rigorous control under the owners of authoritative pages. Thus, spammers design an approach called link spam

Some famous search engines such as Google, Yahoo show a clear attitude to object this disingenuous behavior, and Google have taken actions to punish the websites that use spam tricks, even some famous websites were demoted in the search result rankings because of their spam behavior. However, there are great many other spam sites successfully dodging the detection of search engines, so combating web spam has become one of the major challenges faced by search engines

### 1.1. Links

With link popularity taking on a greater importance in the calculation of relevancy, the spammer's attention has turned to how to manipulate this factor. Link popularity has two components: the authority component (number of links from other resources to this resource) and the hub component (number of links from this resource to other resources).

## 1.2. Hub and Authority in HITS

Authority (n) = ∑hub(m), for all m pointing to n Hub (n) = ∑auth(m), for all m pointed to by n

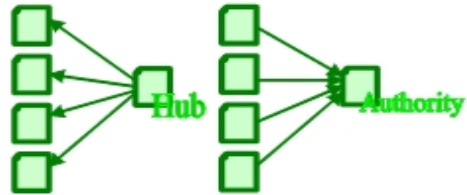

Figure 1. Hubs and Authorities

Two independent efforts in the late 1990 that have profound influence on link analysis were Brin & Page's PageRank and Jon Kleinberg's work on HITS. The details of both algorithms are described in the rest of this section. PageRank and HITS are the two most important ranking approaches in web search; PageRank was used in Google and HITS was extended and applied in AskJeeves. However, modern search engines use not just a single ranking algorithm but a combination of many algorithms and moreover it is not revealed. The simple notation of PageRank is (1).

$$PR(u) = (1 - \alpha) \sum_{v:v \to u} \frac{PR(v)}{O(u)} + \alpha \frac{1}{N}$$

(1)

John Kleinberg proposed [61] that web documents had two important properties, called hubness and authority, as well as a mechanism to calculate them. Pages functioning as good hubs have links pointing to many good authority pages, and good authorities are pages to which many good hubs point. Thus, in his Hyperlink- Induced Topic Search (HITS) approach to broad topic information discovery, the score of a hub (authority) depended on the sum of the scores of the connected authorities (hubs):

$$A(u) = \sum_{v:v \to u} H(v) \qquad H(v) = \sum_{u:u \to v} A(u)$$

(2)

• I(v): in-degree of page v
• O(v): out-degree of page v
• A(v): authority score of page v
• H(v): hub score of page v
• W: the set of web pages
• N: the number of pages in W
• α: the probability of a random jump in the random surfer model
• p ! q: there is a hyperlink on page p that points to q

Techniques such as link farms have been developed to subvert both the authority and hub components.

## 2. RELATED LITERATURE

Many studies on web spam are carried out in previous works. [5] Classified web spam into two categories, one is various techniques to raise the search result rankings, and another is hiding techniques to make the spam activities invisible to users. [3] introduced a method similar to PageRank, first some highly trustworthy sites were selected as seed set, and assigned each of these seeds a initial value, then the values was propagated to other pages in the light of outlink relationship.

After several iterations these pages with high values tend to be non-spam pages. [6] proposed an improved TrustRank algorithm which took both the content quality and link quality into consideration, and this algorithm prevents the inequity that valuable sites mistakenly point to some bad pages will get low scores.

However, the detection of the content quality of web pages is a time consuming problem. [7] introduced the use of topical information to partition the seed set and calculate trust scores for each topic separately, and a combination of these trust scores for a page is used to determine its ranking. [8] focused on how to take full advantage of the information contained in reputable sites, and they adopted an ensemble classification strategy which provides a well-founded mechanism to integrate existing learning algorithm for spam detection. [4] took advantage of the fact that link spam tends to result in drastic changes of links in a short period and proposed using temporal information such as Inlink Growth Rate and Inlink Death Rate in detection of link spam. Clustering the spam is done through a clear means of examining the web structure and it is elaborated in the next part of the paper.

## 3. CLUSTERING THE SPAM AND DBSPAMCLUST

Clustering is a mathematical tool that attempts to discover structures or certain patterns in a data set, where the objects inside each cluster show a certain degree of similarity. Fuzzy clustering allows each feature vector to belong to more than one cluster with different membership degrees (between 0 and 1) and vague or fuzzy boundaries between clusters. The one issue relating to fuzzy clustering is to optimal number of clusters $K$ to be created has to be determined. And the data characterized by large variability's in cluster shape, cluster density, and the number of points (feature vectors) in different clusters have to be handled.

Construct domain based clusters based on the query. For each webpage y there exists a coefficient giving the degree of being in the ith cluster $V_k(y)$. Usually, the sum of those coefficients for any given y is defined to be 1:

$$\forall y \left( \sum_{k=1}^{num.clusters} V_k \left( y \right) = 1 \right) \tag{3}$$

With fuzzy $c$-means, the centroid of a cluster is the mean of all points, weighted by their degree of belonging to the cluster:

$$center_k = \frac{\sum_y v_k \left( y \right)^m y}{\sum_y v_k \left( y \right)^m} \tag{4}$$

The degree of belonging is related to the inverse of the distance to the cluster center:

$$V_k(y) = \frac{1}{Deg(Center_k, y)} \qquad (5)$$

then the coefficients are normalized and fuzzyfied with a real parameter $m > 1$ so that their sum is 1. So

$$V_k(y) = \frac{1}{\sum_j \left( \frac{d(center_k, y)}{d(center_j, y)} \right)^{2/(m-1)}} \qquad (6)$$

For $m$ equal to 2, this is equivalent to normalizing the coefficient linearly to make their sum 1. When $m$ is close to 1, then cluster center closest to the point is given much more weight than the others. Steps involved in creating the WWW$^{DBspamCLUS}$ listed as follows.

1. Choose a number of clusters.

2. Assign randomly to each webpage coefficients for being in the clusters.

3. Repeat until the algorithm has converged (that is, the coefficients' change between two iterations is no more than $\mathcal{E}$, the given sensitivity threshold) :

4. Compute the centroid for each cluster, using the formula above.

5. For each point, compute its coefficients of being in the clusters, using the formula above. The algorithm minimizes intra-cluster variance as well.

Table 1 Classifier Formulation

| | | Prediction | |
| --- | --- | --- | --- |
| | | Non-spam | Spam |
| True label | Non-Spam | x | y |
| | Spam | z | w |

True positive rate TPR $= \dfrac{w}{z+w}$

False positive rate FPR $= \dfrac{y}{y+x}$

F-Measure P $= \dfrac{w}{y+w}$

F $= 2 \; \dfrac{TPR \cdot FPR}{TPR+FPR}$

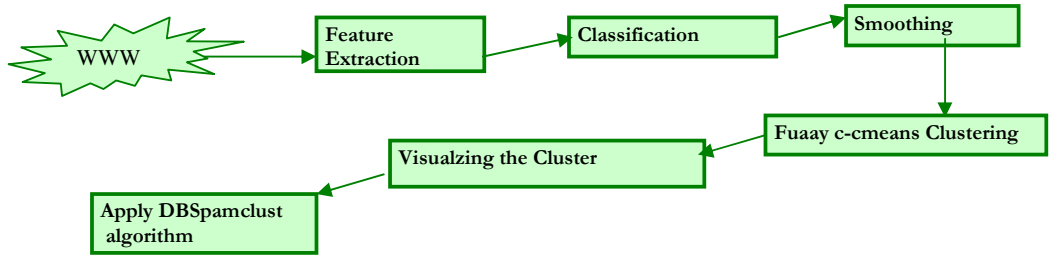

Figure 2 Methodology of DBSpamClust

For extracting the features the first data set is obtained by using web crawler. For each page, links and its contents are obtained. From data set, a full graph is built. For each host and page, certain features are computed. Link-based features are extracted from host graph. Content-based feature are extracted from individual pages.Some important link-based features are listed:

- Degree-related measures
- PageRank
- Estimation of supporters – Authority
- Network path length
- Reciprocal links

The base classifier from link-based content-based features has to be created. And apply cost-sensitive decision tree to classify spam and non-spam hosts.

Table 2 Decision Tree Formulation Parameters

| Cost ratio | 1 | 10 | 20 | 30 | 50 |
|---|---|---|---|---|---|
| True positive rate | 65.0% | 67.0% | 74.6% | 79.1% | 86.0% |
| False positive rate | 4.6% | 7.8% | 9.5% | 11.7% | 14.4% |
| **F-Measure** | 0.632 | 0.633 | 0.646 | 0.642 | 0.594 |

As such mentioned in the parameters in the table 2 the decision tree could be constrained to capture the spam. Based on the classifier the hosts are labelled as either spam or non-spam. Now the Spam tends to be clustered on the Web, whereas non-spam nodes tend to be linked by very few spam nodes, and usually link to no spam nodes. Spam nodes are mainly linked by spam nodes. This principle is based on the assumption that good pages seldom point to good pages. Now consider w, URL of the sample webpage it resides in a domain which was treated as a cluster CLUS(w). Now consider $^{IN}$(CLUS(w)), the incoming links to the particular domain of w. Also assume that the link concludes at certain point in the domain and lead to another domain and called as CON$^{Temp}$. The outgoing links which leads to the different cluster can be considered as $^{OUT}$(CLUS(w)). TRA$^{LVL}$ could be set to a fixed value to restrict the iteration. Consider the threshold TV$^{DSCLUS}$. Now if the web page exceeds the threshold limit then the page will be marked as spam page.

1.  For each URL x in $^{IN}$(CLUS(w)) perform

2.  If CLUS(x) != CLUS(w) and not present in $^{IN}$(CLUS(w)), then add it in the Cluster

3.  Set w as CON$^{Temp}$ and set current level of traversal TRA$^{CUR}$ to 0.

4.  If level, TRA$^{CUR}$<= TRA$^{LVL}$ then, For each URL y in $^{OUT}$(CLUS(w)) perform

    a. If CLUS(y) != CLUS(x) and if it is not found in $^{OUT}$(CLUS(w)) then add it to the domain list of outgoing links.

    b. Else if  CLUS(y) = = CLUS(x), then set TRA$^{CUR}$++ and set y as TRA$^{LVL}$ and repeat step 3 and 4.

    c. Calculate the average path length for each website using its traversal depth.

    d. Analyze the degree distribution. For a vertex $i$ , the degree $Ki$ is an important property and denotes the number of its nearest neighbors or the number of its edges. For a network, the essential property is the degree distribution $P(K)$, which expresses the probability that a randomly selected vertex has exactly $K$ edges.

    e. Analyze the reciprocal links which may predict the presence of the bi-partite graph which indicates the strong presence of the spam and may help in reducing the false negatives.

5.  Calculate the intersection of   $^{IN}$(CLUS(w)) and $^{OUT}$(CLUS(w)). If the number of elements in the intersection set is equal to or bigger than the threshold TV$^{DSCLUS}$, mark x as a bad page.

6.  Repeat the steps for every search result URL, x.

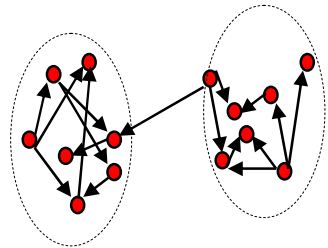

Figure 3 Clusters to be formulated based on domains

The clustered community is derived with the help of the algorithm and the new community formation could be done to locate the spam and non spam communities as such illustrated in figure 3. Since only few domains are handled in this paper for initiative process and the calculated results shows that the spam identification could be done in efficient manner for the considered data set

Table 3 Resulting Clusters

|  | *Base* | *Cluster* |
|---|---|---|
| Without grouping |  |  |
| TPR | 71.6% | 70.5% |
| FPR | 12.5% | 11.8% |
| F-Measure | 0.646 | 0.643 |
| With grouping |  |  |
| TPR | 77.7% | 75.9% |
| FPR | 6.7% | 6.0% |
| **F-Measure** | **0.723** | **0.728** |

When clustering the false positive identification (ie.) the spam could be derivate easily whereas the performance increases and it is measured in terms of precision and recall as mentioned earlier. One interesting observation noticed when crawling the web for the substantiating the efficiency of the proposed DBSpamClust, an experiment was conducted using the search results obtained from Yahoo, Google, Ask and Altavista Search Engine reveals that the complex queries face less number of spam and simple, commercial, sexually implicit queries is spammed a lot which posess less accuracy towards the query as such mentioned in figure 4.

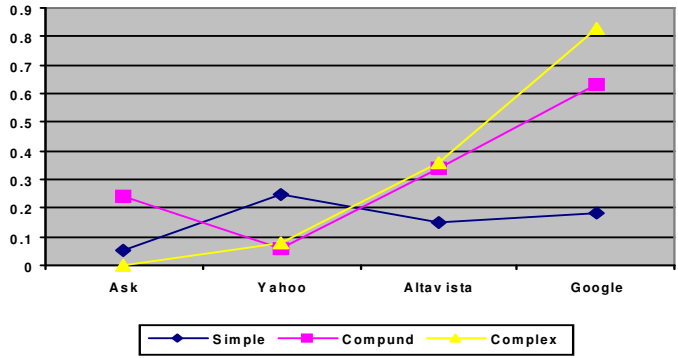

Figure 4 Query nature Vs spamness

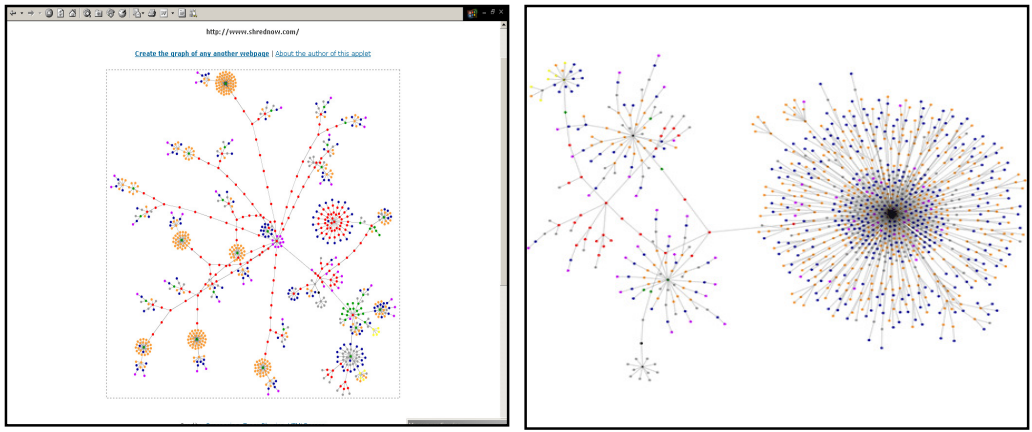

Figure 5 Simulated result for Spam clusters

The graph cluster shown in the figure 5 illustrates that the spam cluster possesses a structure which violates the power law distribution and reviewing the topological structure of the website may reveal more interesting findings in spamdexing detection. Figure 6 shows an extremely spammed link webpage while crawling the data for the query "online earning in tamilnadu", this website also possesses the bipartite graph which leads to enormous link inculcated for rank merit. It achieves the rank within top 5 results consistently for timestamp of Sep'10 to Nov'10(As such monitored).

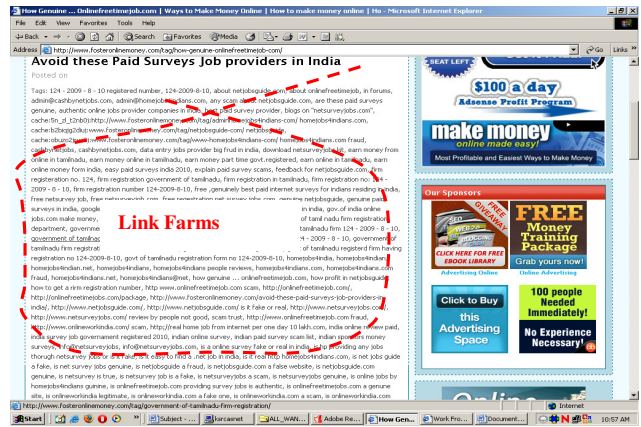

Figure 6 Web snapshot of Link farm building

Table 4 Some Observations on spam pages

| Concealed spam pages identified – for query relating to "online earning in tamilnadu" | | | | |
|---|---|---|---|---|
| Domain name | Description | Inter sections | Classification | Connection type |
| http://www.fosteronlinemoney.com | Yahoo detects 426 back links and 117 pages where Google detects 13 back links and 1330 pages (Listed at top 5 in Google) | 5 | Link Farm(Accumulated Links – seems to be Reciprocal) | Fully connected |
| **http://india.newads.org** | Yahoo detects 16947 back links and 516 pages and Google detects 11 back links and 18400 pages and Clusty detects 19800 back links | 7 | Boosting Links (One page possesses multi-link pointing which boost the ranking) | Fully connected |

## 4. CONCLUSION

As it is stated in the graph in figure 9 the complex queries faces less number of spam whereas the simple queries subject to spamcity massively. It is not necessary to consider how difficult a spamming technique could be almost all of them are designed to have the effect of manipulating the factors that are used, or believed to be used, by popular search engines in their ranking algorithms. This paper presented the DBSpamClust algorithm is to identify link farm spam pages in the search engines' results. As certain outgoing spam links are intentionally hidden by spammers, some of these pages would be able to bypass the earlier algorithms.

Hence, DBSpamClust is suggested to improve the efficiency in link spam detection by analyzing additional link farms based on some constraints as such motioned in the algorithm. The experiment conducted had also proven that additional potential spam pages could be identified using DBSpamClust. One of the possible improvements is by integrating the weight of web page content relevancy into DBSpamClust and formulating a collaborative constraint based filter.

# REFERECES

## Authors

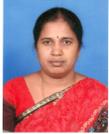 Dr.S.K.Jayanthi received the M.Sc., M.Phil., PGDCA, Ph.D in Computer Science from Bharathiar University in 1987, 1988, 1996 and 2007 respectively. She is currently working as an Associate Professor, Head of the Department of Computer Science in Vellalar College for Women. She secured District First Rank in SSLC under Backward Community. Her research interest includes Image Processing, Pattern Recognition and Fuzzy Systems. She has guided 18 M.Phil Scholars and currently 4 M.Phil Scholars and 4 Ph.D Scholars are pursuing their degree under her supervision. She is a member of ISTE, IEEE and Life Member of Indian Science Congress. She has published 2 papers in International Journals and one paper in National Journal and published an article in Reputed Book. She has presented 10 papers in International level Conferences/Seminars, 14 papers in National level Conferences/Seminars and participated in around 25 Workshops/Seminars/Conferences/FDP.

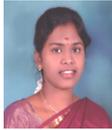 S.Sasikala, currently working as a Lecturer in K.S.R. College of Arts & Science has received the B.Sc(CS) from the Bharathiar University, M.Sc(CS) from the Periyar University, M.C.A. from Periyar University , M.Phil from Periyar University, PGDPM & IR from Alagappa university in 2001, 2003, 2006, 2008 and 2009 respectively. And she is currently pursuing her Ph.D in computer science at Bharathiar University. Her area of Doctoral research is Web mining. She secured University First Rank in M.Sc(CS) Programme under Periyar University and received Gold Medal from Tamilnadu State Governor Dr.RamMohanRao in 2004. She has published 1 paper in international journal of Advances in Computational Sciences and Technology (ACST). She has presented 8 papers in International Conferences/Seminars, 17 papers in National Conferences/Seminars and participated in 4 National Conferences/Seminars and 2 Workshops.